\documentclass[journal,12pt,onecolumn,letterpaper]{IEEEtran}
\usepackage{arxiv}

\usepackage[utf8]{inputenc} 
\usepackage[T1]{fontenc}    
\usepackage{hyperref}       
\usepackage{url}            
\usepackage{booktabs}       
\usepackage{amsfonts}       
\usepackage{nicefrac}       
\usepackage{microtype}      
\usepackage{lipsum}
\usepackage{amsmath}
\usepackage{amssymb}
\usepackage{lscape}
\usepackage{color}
\usepackage{array}
\usepackage{rotating}
\usepackage{adjustbox}
\usepackage{tabularray}
\usepackage{graphicx}

\title{MagicEye: An Intelligent Wearable Towards Independent Living of Visually Impaired}

\author{
 Sibi C. Sethuraman\\
  Centre of Excellence, Artificial Intelligence \& Robotics (AIR),\\
  School of Computer Science and Engineering\\
  VIT-AP University, India \\
  \texttt{chakkaravarthy.sibi@vitap.ac.in} \\
   \And
Gaurav R. Tadkapally \\
 Centre of Excellence, Artificial Intelligence \& Robotics (AIR),\\
  School of Computer Science and Engineering\\
  VIT-AP University, India \\
  \texttt{gauravreddy008@gmail.com} \\
 \And
 Saraju P. Mohanty\\
Department of Computer Science and Engineering,\\
University of North Texas, \\
TX 76207, USA, \\
  \texttt{saraju.mohanty@unt.edu}\\
 \And
  Gautam Galada \\
  Centre of Excellence, Artificial Intelligence \& Robotics (AIR),\\
  School of Computer Science and Engineering\\
  VIT-AP University, India \\
   \And
  Anitha Subramanian \\
  Centre of Excellence, Artificial Intelligence \& Robotics (AIR),\\
  School of Computer Science and Engineering\\
  VIT-AP University, India \\
}

\begin{document}

\maketitle

\begin{abstract}
Individuals with visual impairments often face a multitude of challenging obstacles in their daily lives. Vision impairment can severely impair a person's ability to work, navigate, and retain independence. This can result in educational limits, a higher risk of accidents, and a plethora of other issues. To address these challenges, we present MagicEye, a state-of-the-art intelligent wearable device designed to assist visually impaired individuals. MagicEye employs a custom-trained CNN-based object detection model, capable of recognizing a wide range of indoor and outdoor objects frequently encountered in daily life. With a total of 35 classes, the neural network employed by MagicEye has been specifically designed to achieve high levels of efficiency and precision in object detection. The device is also equipped with facial recognition and currency identification modules, providing invaluable assistance to the visually impaired.  In addition, MagicEye features a GPS sensor for navigation, allowing users to move about with ease, as well as a proximity sensor for detecting nearby objects without physical contact. In summary, MagicEye is an innovative and highly advanced wearable device that has been designed to address the many challenges faced by individuals with visual impairments. It is equipped with state-of-the-art object detection and navigation capabilities that are tailored to the needs of the visually impaired, making it one of the most promising solutions to assist those who are struggling with visual impairments.

\end{abstract}


\section{Introduction}
\label{Sec:Introduction}

It is estimated that over 2.2 billion people have vision impairment, among which 1.4 million children under the age of 15 are completely visually impaired \cite{pascolini_global_2012}. Although most people over the age of 50 years encounter vision loss, it is likely to pose a risk to all age groups \cite{ulldemolins_social_2012}. Vision impairment causes impaired intellectual and social development of the individual, thus affecting numerous aspects of life.  visual impairment is considered one of the top disabilities among adults and a frightening disorder in children \cite{rahi_severe_2003}. Unlike other conditions, vision impairment can be caused by ageing and uncontrolled diabetes, which is a prevalent health condition in this day and age. Furthermore,  visual impairment can also be caused by genetic mutations or congenital disabilities, resulting in the person being born with the disability \cite{dineen_causes_2007}. Likewise, the child is restricted to growing in unusual environments compared to others, consequently affecting their growth.

\begin{figure*}[htbp]
\centering
\includegraphics[width=0.85\textwidth]{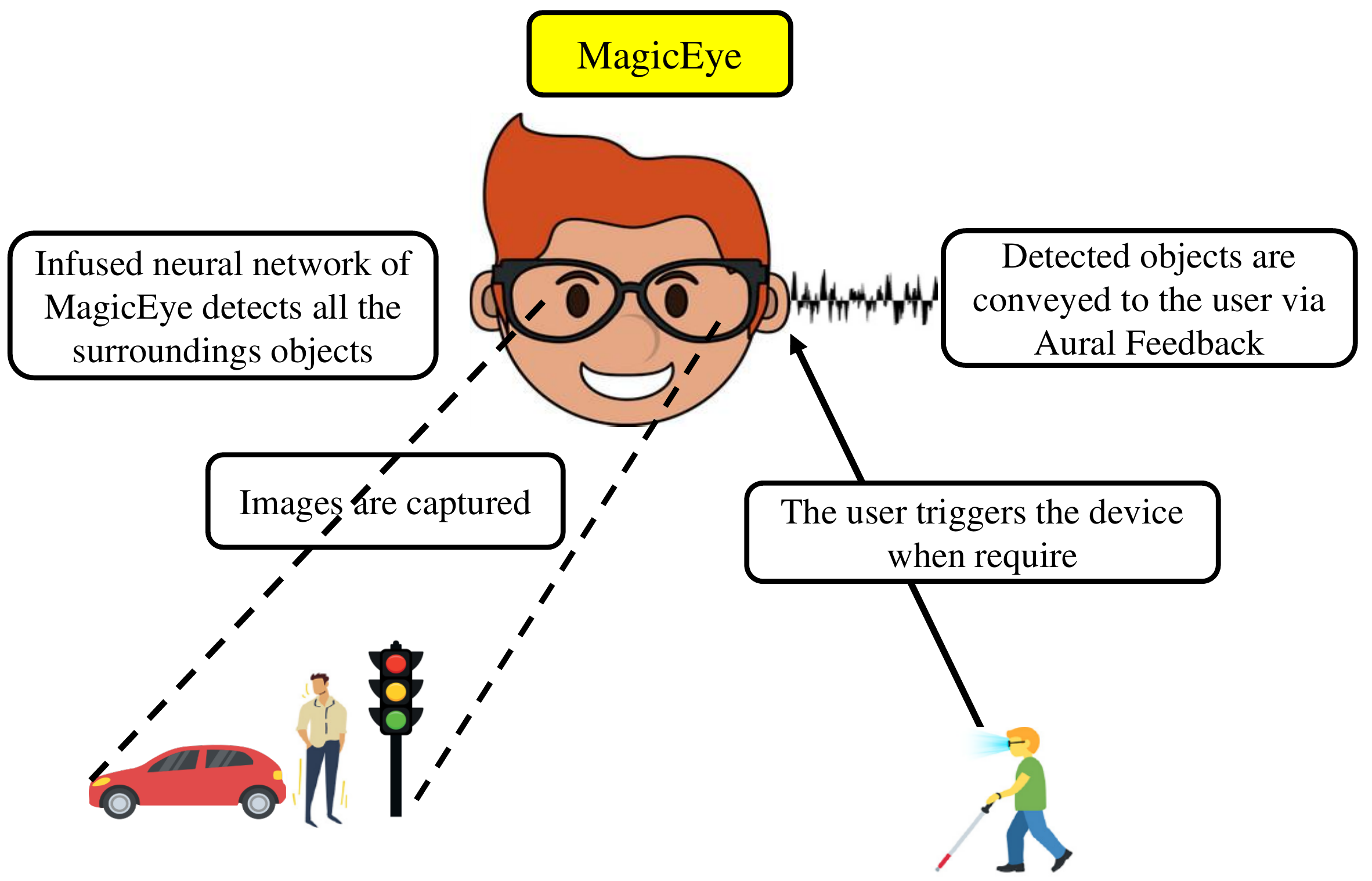}
\caption{A conceptual overview of the proposed solution}
\label{FIG:Conceptual_overview_proposed_solution}
\end{figure*}

Visually impaired people require demanding efforts to carry out their daily tasks. Simple chores, such as walking across the street or cooking, become highly risky. To navigate the city, an individual suffering from vision loss needs to have a clear mental image of every street and memorise the surroundings very quickly; otherwise, it is highly unsafe and implausible. A survey shows that nearly all people have had accidents due to visual loss \cite{frick_economic_2007}. Falling into unclosed pits, falling from stairs, and bumping into other obstacles on the sidewalks are a few different examples of these occurrences. Even if an individual decides to limit themselves indoors, they still need to communicate, read, write, continuously recognise their surroundings and objects and perform other essential tasks, which require a great effort to be carried out. In conclusion, unlike others, it is challenging for the individual to lead a normal lifestyle. 

Despite the struggles, specialised organisations and governments have established laws and standards to support people with visual impairment. In addition, countless adequate navigation supports and other aids help these individuals with their daily activities. One of the popular techniques: Braille, helps the visually challenged identify objects and perceive bits of information \cite{jimenez_biography_2009}. It is a form of written language specialised for the abled individuals, in which the characters are patterns of raised dots such that the person can quickly feel the letters with their fingertips and comprehend them. However, the braille literacy rate is meagre compared to regular literacy.

Furthermore, there are other traditional aids, such as a mobility cane, a stick used to navigate and avoid barriers, and guide dogs trained to assist the visually impaired and help them navigate around places. Some works target to improve quality of life through use of smart sticks, but don't target visually impaired population \cite{Rachakonda_IFIP-IoT_2021, Rachakonda_IFIP-IoT_2019}. Nevertheless, these do not eliminate the difficulties of the people. For example, instinctively leading their way by solely relying on a mobility cane can be dangerous. It is ineffective with distant obstacles, and they must warn the individual about hazardous environments, such as traffic \cite{calder_ecological_2010}. Likewise, many smartphone software or applications have been introduced to help the visually impaired navigate using voice-based services. Nevertheless, their proper use and integration still need to be improved and insufficient. 

Rest of this paper is organized in the following manner: Section \ref{Sec:Novel_Contributions} summarizes the novel contributions of this work. Section \ref{Sec:Prior_Research} presents the existing works on this area of research. Section \ref{Sec:Proposed_Framework} discusses the details of the proposed MagicEye. Section \ref{Sec:Experimental_Results} has the discussions of the validations and experiments of this work. Section \ref{Sec:Conclusion} presents the research findings of this work. Section \ref{Sec:Future_Research} presents some discussions on the possible future directions of this work.

\section{Novel Contributions of the Current Paper}
\label{Sec:Novel_Contributions}

\subsection{Problem Addressed}
Recent advancements in the fields of the Internet of Things (IoT) and Artificial Intelligence (AI) have led to the development of several visual aids for able individuals, aimed at providing convenience and facilitating recognition and detection of surroundings. However, current solutions have fallen short in addressing the needs of visually impaired individuals. For example, some researchers have attempted to improve mobility canes with the integration of proximity sensors, thus alerting users to obstacles in their path. But such proposals are inherently limited and lack reliability in unpredictable environments, such as while crossing roads, where the individual must have full situational awareness. In light of these limitations, it is imperative to introduce a more sophisticated and comprehensive solution to address the unique challenges faced by visually impaired individuals. The primary objective of this study is to address these challenges and provide a comprehensive solution to assist visually impaired individuals. The current study focuses on the following key challenges:
\begin{itemize}
\item \textbf{Obstacle Detection}: Navigating unfamiliar environments can be especially challenging for visually impaired individuals. In many cases, they must rely on touch, sound, or a cane to detect obstacles in their path.

\item \textbf{Peer Recognition}: Visually impaired individuals often struggle to recognize their peers, which can be especially challenging in social settings. 

\item \textbf{Currency Recognition}: The visually impaired also face challenges in recognizing and handling currency. The proposed solutions need to incorporate algorithms that enables the user to easily recognize and handle banknotes, improving their independence and ease of use in day-to-day transactions.
\end{itemize}

\subsection{Solution Proposed in the Current Paper}

In this study, we introduce MagicEye, a highly sophisticated and innovative wearable device that provides real-time obstacle detection and recognition for visually impaired individuals. This cutting-edge technology has been designed to be used in conjunction with GPS and proximity sensors, offering turn-by-turn navigation guidance through audio feedback while simultaneously alerting the user of any imminent hazards. Additionally, the device integrates state-of-the-art facial recognition architectures, enabling the user to quickly and accurately identify their peers. Furthermore, a currency classification algorithm has been seamlessly integrated into the system, offering individuals with visual impairments the ability to recognize banknotes with ease and facilitating a wider range of convenient, hand-to-hand transactions. 

\subsection{Significance of the Proposed Solution}

The significance of the proposed solution, MagicEye, is two-fold. Firstly, it provides a comprehensive solution for the visually impaired individuals, which is currently missing in the existing solutions. Unlike the traditional devices equipped with proximity sensors, MagicEye offers real-time turn-by-turn navigation with audio feedback, along with the detection and recognition of obstacles, thereby increasing the safety and mobility of the user. Secondly, the system's integration of state-of-the-art facial recognition and currency classification algorithms further enhances the independence and convenience of the visually impaired. The system's facial recognition capabilities allow the user to identify their peers in real-time, which can be particularly useful in various social and professional scenarios. Further on, the facial recognition model goes beyond basic identification by also providing information on the individual's gender, race, age, and facial expression. The currency classification algorithm, on the other hand, empowers the visually impaired to independently perform hand-to-hand transactions, eliminating the need for assistance in counting or verifying banknotes. 

MagicEye represents a significant advancement in the field of AI-based solutions for the visually impaired. The system's robust and comprehensive features have the potential to significantly improve the quality of life of visually impaired individuals and increase their independence in daily life. As demonstrated in Fig. \ref{FIG:Conceptual_overview_proposed_solution}, the conceptual framework of MagicEye presents a holistic and sophisticated solution to the challenges faced by visually impaired individuals.






\begin{sidewaystable}[htbp]
\centering
\caption{Comparison of MagicEye with similar assistive systems}
\label{TBL:Product_Comparison}
\resizebox{\linewidth}{!}{%
\begin{tblr}{
  width = \linewidth,
  colspec = {Q[110]Q[156]Q[310]Q[54]Q[87]Q[142]Q[71]},
  row{1} = {c},
  cell{1-11}{1-7} = {c},
  cell{1-11}{3} = {l},
  hline{1-2,12} = {-}{},
}
Assistive System & Methodology Employed & Findings and Outcome & Precision & Indoor/Outdoor & Facial and Currency recognition modules & GPS Navigation\\

Garrido et al. (2012) \cite{garrido_support_2012}  & Near Field Communication (NFC) tags & {\labelitemi\hspace{\dimexpr\labelsep+0.5\tabcolsep}High accurate predictions.\\\labelitemi\hspace{\dimexpr\labelsep+0.5\tabcolsep}Extremely restricted operational distance.} & Low & Indoors & No & No\\

Nagarajan et al. (2020) \cite{nagarajan_localization_2020} & Bluetooth Low Energy~(BLE) & {\labelitemi\hspace{\dimexpr\labelsep+0.5\tabcolsep}Relies on BLE beacons distributed throughout the environment\\\labelitemi\hspace{\dimexpr\labelsep+0.5\tabcolsep}Operation is contingent on proximity to these beacons} & Medium & Indoors & No & No\\

SUGAR (2015) \cite{martinez-sala_design_2015}& Ultra-Wideband~(UWB) & \labelitemi\hspace{\dimexpr\labelsep+0.5\tabcolsep}Suitable only for static environments. & Medium & Indoors & No & No\\

ARIANNA (2016) \cite{croce_enhancing_2016} & Smartphone camera & {\labelitemi\hspace{\dimexpr\labelsep+0.5\tabcolsep}Relies on colored floor strips for path identification,\\\labelitemi\hspace{\dimexpr\labelsep+0.5\tabcolsep}Can only accommodate a constant user speed.} & Good & Indoors & No & No\\

NavCog3 (2019) \cite{sato_navcog3_2019} & Bluetooth Low Energy~(BLE)  & {\labelitemi\hspace{\dimexpr\labelsep+0.5\tabcolsep}BLE beacons and geolocation tracking are used in combination\\\labelitemi\hspace{\dimexpr\labelsep+0.5\tabcolsep}The system does not provide support for object or scene recognition} & Good & Indoors & No & Yes\\

M. A. Khan et al. (2020) \cite{khan_ai-based_2020} & Deep CNN (MobileNet) & {\labelitemi\hspace{\dimexpr\labelsep+0.5\tabcolsep}Dual capabilities: object detection and distance measurement\\\labelitemi\hspace{\dimexpr\labelsep+0.5\tabcolsep}Unreliable in complex outdoor environments} & Good & Both & No & No\\

Chang et al. (2021) \cite{9253595} & Deep CNN (InceptionNet) & {\labelitemi\hspace{\dimexpr\labelsep+0.5\tabcolsep}Employs fall detection and aerial obstacle avoidance pipeline\\\labelitemi\hspace{\dimexpr\labelsep+0.5\tabcolsep}Lacks support for object or scene recognition} & Excellent & Very limited outdoors & No & No\\

R. C. Joshi et al. (2020) \cite{joshi_yolo-v3_2020} & Deep CNN (YoloV3) & {\labelitemi\hspace{\dimexpr\labelsep+0.5\tabcolsep}95.71\% average detection and 100\% average recognition rate achieved\\\labelitemi\hspace{\dimexpr\labelsep+0.5\tabcolsep}Limited to detecting banknotes} & Excellent & Limited outdoors & No/Yes & No\\

M. Affif et al. (2020) \cite{afif_evaluation_2020} & Deep CNN (RetinaNet) & {\labelitemi\hspace{\dimexpr\labelsep+0.5\tabcolsep}High precision in detection\\\labelitemi\hspace{\dimexpr\labelsep+0.5\tabcolsep}Limited to recognizing only 16 subjects} & Excellent & Limited outdoors & No & No\\

\textbf{MagicEye (Current Paper)} & Deep CNN (YoloV5) & {\labelitemi\hspace{\dimexpr\labelsep+0.5\tabcolsep}Swiftly detects objects in real-time with high precision\\\labelitemi\hspace{\dimexpr\labelsep+0.5\tabcolsep}Outperforms other works by accurately detecting over 35 subjects\\\labelitemi\hspace{\dimexpr\labelsep+0.5\tabcolsep}Includes facial recognition and currency detection for a more intuitive user experience.} & Excellent & Both & Yes & Yes

\end{tblr}
}
\end{sidewaystable}

\section{Prior Related Research}
\label{Sec:Prior_Research}

In recent years, the rapid pace of technological advancements has led to the development of a wide range of assistive technologies aimed at aiding the visually impaired in navigating unfamiliar environments \cite{real_navigation_2019}. The primary objective of such technologies is to enable visually impaired individuals to detect obstacles and recognize their surroundings. However, a few systems have also introduced capabilities for accessing printed materials and social interaction.

The field of assistive technologies for the visually impaired has garnered significant research interest over the past decade, resulting in the development of a variety of effective methods, including wearable devices, intelligent canes, smartphone-based applications, and retinal implants. Furthermore, research in this field has expanded to include the development of neural brain implants that could potentially cure vision loss \cite{theogarajan_strategies_2012}. The result of these advancements is a diverse range of solutions that can cater to the specific needs and preferences of visually impaired individuals, enabling them to live more independent and fulfilling lives.

\subsection{Sensory-based systems}

The integration of various technologies and sensors has led to the development of numerous non-vision-based solutions \cite{khan_analysis_2021}. While these systems can be useful for visually challenged individuals in solving the localization problem, they cannot be standalone solutions.

Currently available smartphones are equipped with built-in Near-Field Communication (NFC) technology, which has been utilized in a few proposed methods for indoor navigation using NFC tags \cite{garrido_support_2012}. Although this approach provides high accuracy, its limited operational range of only a few centimeters makes it unreliable over longer distances. In contrast, some systems have utilized Bluetooth Low Energy (BLE), another wireless technology, to establish a network of navigational tags \cite{nagarajan_localization_2020}. BLE is more reliable over longer distances than NFC, thus requiring fewer tags for a given area. However, it only accurately detects the user's location at specific points and cannot continuously monitor the user. Ultra-Wideband (UWB) technology has also been utilized in a few visual aid systems \cite{alnafessah_developing_2016}. UWB's operational range of 90 meters and relatively high accuracy make it well-suited for use in larger buildings. This technology was also used in the implementation of the "SUGAR" system, which helps visually impaired individuals navigate indoors \cite{martinez-sala_design_2015}. Additionally, non-vision-based solutions utilizing various other wireless technologies such as Infrared (IR) \cite{bendanillo_sight-man_2020} and Wi-Fi have also been introduced. However, such systems are limited to functioning only indoors and primarily serve to detect obstacles rather than recognize them.

\subsection{Vision-based systems}

Vision-based systems typically identify obstacles, which gives them an advantage over non-vision-based alternatives. Vision-based systems are intended to detect, identify, and inform the user of their surroundings via haptic or auditory feedback. Using RGB-D cameras, numerous studies have developed wearable navigation devices for able-bodied individuals \cite{kayukawa_blindpilot_2020, zhang_slam_2015}. Likewise, a few systems have adapted smartphone cameras and developed multiple software applications to aid users in navigating various locations \cite{khan_insight_2021}. One is ARIANNA, a smartphone-based navigation system that uses a smartphone's camera to navigate through locations; the user receives tactile feedback \cite{croce_arianna_2014,croce_enhancing_2016}. NavCog3 is a similar smartphone-based application but further equipped with off-the-shelf Bluetooth beacons installed in surrounding environments \cite{sato_navcog3_2019}. Similarly, a cloud-based system was developed utilizing the smartphone’s GPS and camera \cite{lapyko_cloud-based_2014}. Although considerable efforts have been made to implement these systems, there is room for improvement. It is implausible for a visually impaired individual to effectively traverse their environment using only a smartphone. While the majority of these systems are restricted to indoor environments, as they require wireless sensors to be installed in the surrounding environments of a structure, real-time tracking of the user's location requires a wireless network map.

Notably, our proposed system MagicEye employs the equipped deep neural network to detect and recognise a visually impaired individual's environments. Comparatively, numerous methods employing the same architecture as MagicEye have been proposed. In \cite{afif_evaluation_2020}, the authors propose an assistive navigation system, leveraging a Deep Convolutional Neural Network (DCNN), famously known as RetinaNet. Although the authors have achieved commendable results, the proposed system is confined to being used indoors. Another study followed a similar approach and utilized the MobileNet architecture, a popular DCNN, to detect various objects and animals \cite{khan_ai-based_2020}. Essentially, the aforementioned systems utilize Region-Based Object Detection algorithms, commonly referred to as RCNNs. Despite their high accuracy and precision, RCNNs require a substantial amount of computational power, making them unreliable for real-time use. Another article presents a wearable assistive system for visually impaired individuals to safely navigate zebra crossings \cite{9253595}. The system includes smart sunglasses, a waist-mounted device, and an intelligent walking cane, which work together using AI edge computing to recognize zebra crossings in real-time. However, the system does not address the larger issue of navigating an unfamiliar environment beyond the zebra crossing, which is where MagicEye aims to help.

On the contrary, a study compared their proposed system to a few other neural network architectures and concluded with the conclusion that Single Shot Detector performed the best \cite{dascalu_usability_2017}. Despite numerous research and advancements, the Single Shot Detector need more characteristics, like swift real-time working or their ability to function in various environments. Furthermore, a few more studies have used a similar architecture with inferior features to MagicEye \cite{shahira_obstacle_2019-1, joshi_yolo-v3_2020}. Although these similar architectures achieved high real-time performance, they failed to address the fundamental and rudimentary aspects of developing intelligent assistive technology for the unsighted \cite{shahira_obstacle_2019-1, joshi_yolo-v3_2020}.

Previous works have established and devised numerous baseline assistive technologies for the unsighted by utilizing multiple technologies and following various pipelines. However, there is still room for improvement, as previous works fail to address several prominent and fundamental issues faced by the visually impaired when using these proposed frameworks. Specifically, the only goal of existing frameworks was to detect and recognize objects in a limited environment, whereas our proposed system, MagicEye, is far more capable. In this paper, we made every effort to address all of the shortcomings introduced in previous research while also developing a novel intelligent wearable system for the visually challenged.


\section{Proposed Framework of the MagicEye}
\label{Sec:Proposed_Framework}

\begin{figure*}[htbp]
\centering
\includegraphics[width=0.95\textwidth]{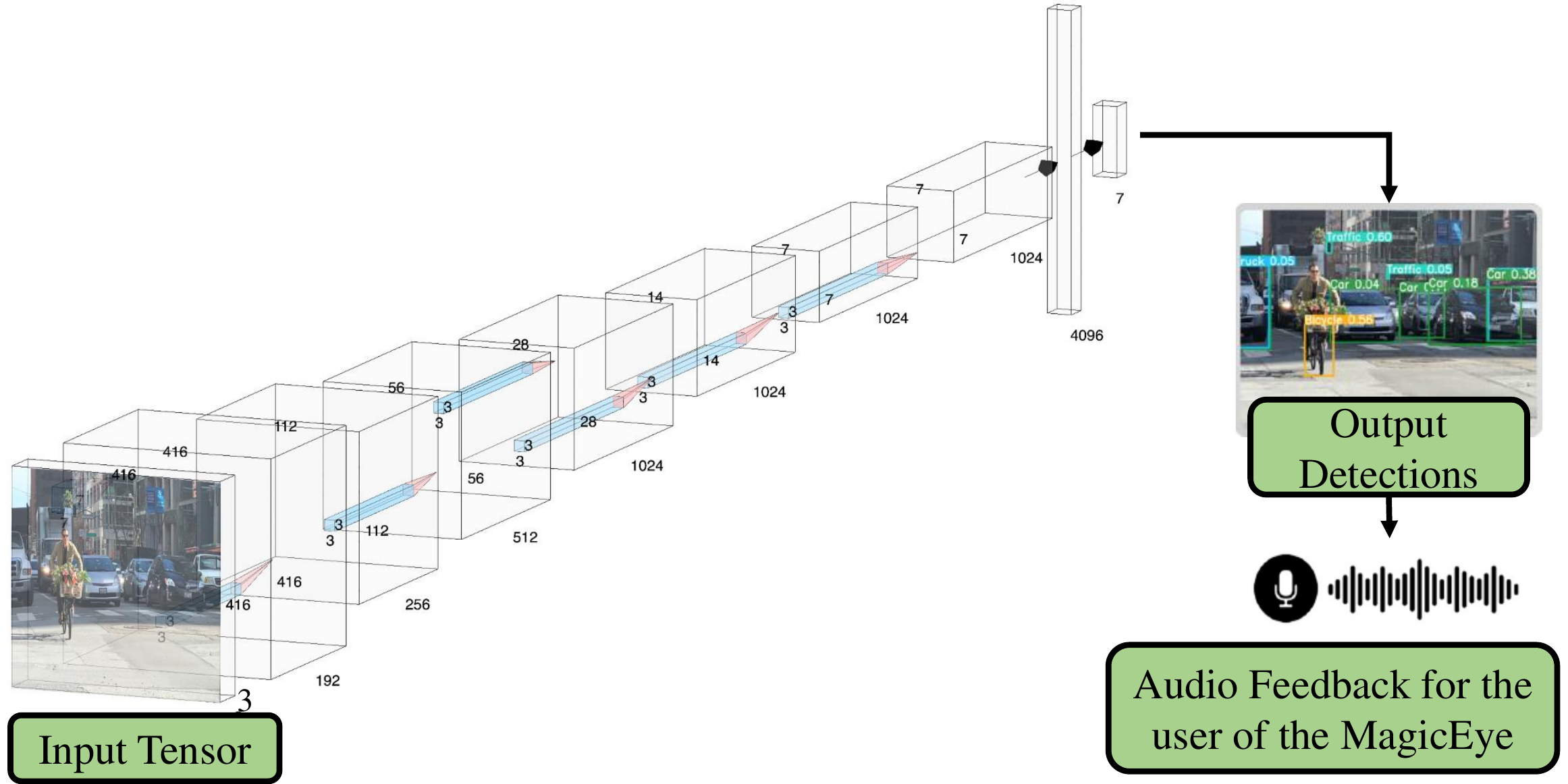}
\caption{The Data Flow is the MagicEye.}
\label{FIG:DataFlow_of_MagicEye}
\end{figure*}

The MagicEye system detects and identifies objects using convolution neural networks as it's core and elemental component. In a problem of machine perception, the built-in neural engine extracts image features that determine the subject's presence and contribute to the subject's identification. The primary objective of existing solutions has always been to detect and identify objects and obstacles in the immediate vicinity of a person. Many researchers have attempted to create the "ideal" assistive system for disabled individuals, many have neglected to address the numerous problems and drawbacks associated with using these systems. In addition, previous proposed solutions are limited to particular environments, primarily indoors. More specifically, the results published in \cite{afif_evaluation_2020, khan_ai-based_2020, joshi_yolo-v3_2020, dascalu_usability_2017, shahira_obstacle_2019-1} can only recognize a limited number of objects, making them unreliable in a real-time scenario. As previously mentioned, the system primarily employs the integration of three distinct modules to yield a viable system. Furthermore, MagicEye's neural network has been trained to recognize 35 distinct classes.. These classes contain objects that are encountered in a person's daily life. As a result, we believe, MagicEye is more reliable than existing solutions. The simplified working of MagicEye is illustrated in Fig. \ref{FIG:DataFlow_of_MagicEye}. The core components of MagicEye are discussed in the remaining of this Section.

\subsection{The Integrated Object Detection System}

In various computer vision applications, assistive technologies employing convolutional neural networks (CNN) have demonstrated state-of-the-art performance \cite{real_navigation_2019}. Specifically, a conventional image classification model employs hundreds of features extraction layers to generate and recognise patterns in an image. In our case, however, our model had to be capable of detecting multiple objects in a single image along with their localization coordinates. 

In a real setting, when a user interacts with our device or the proximity sensor triggers it, the built-in camera captures an image of the user's surroundings. The image is then processed by the neural network equipped by MagicEye, which generates predictions based on the image. Notably, the model first divides the input image into a grid of cells, each of which is responsible for predicting a bounding box if the centre of a bounding box falls within the cell. Likewise, the system would identify and generate predictions for the perceived objects in the image. Finally, the  user is notified of these predictions via Audio Feedback. In this way, despite their vision loss, the individual can be aware of their surroundings at all costs. A generalized working of MagicEye’s infused neural net is represented in Fig. \ref{FIG:Conceptual_overview_proposed_solution}.



Stochastic Gradient Descent (SGD), a generic optimization algorithm, is used as the model optimizer, which helps to tune the model while the training continues \cite{amir_sgd_2021}. Equation \ref{EQN:SGD} describes the workings of a conventional SGD algorithm. However, SGD with momentum, an improved version of the traditional algorithm, is used for MagicEye. Equation \ref{EQN:SGD} illustrates the mathematical equation the model's optimizer uses to update weights for each image sample:
\begin{eqnarray}
w & = & w - \eta \triangledown Q_{i}(w) \\
\label{EQN:SGD}
%
w & = & w -\eta \triangledown Q_{i}(w)+\alpha\triangle w 
\label{EQN:SGD-model}
\end{eqnarray}
In the above expressions, $w$ denotes the weight calculated and $\eta$ represents the learning rate (also called step size), and $\alpha$ represents the exponential decay factor ranging from 0 to 1, which determines the relative contribution of the current gradient to the weight change. $Q_{i}$  denotes empirical risk and is calculated using Equation \ref{EQN:grad}, where it represents the loss function value of the $i$-th image sample:
\begin{equation}
Q( w) \ =\ \frac{1}{n} \ \sum _{i=1}^{n} Q_{i} \ ( w) 
\label{EQN:grad}
\end{equation}

The images are pre-processed, downsized to 416 × 416 pixels, and then aggregated into 32 batches before being transmitted via the network. Additionally, the training pipeline analyses the model for each iteration using the box, object, and classification loss functions.

\subsection{Neural Network Architecture}

\begin{figure*}[htbp]
\centering
\includegraphics[width=\textwidth]{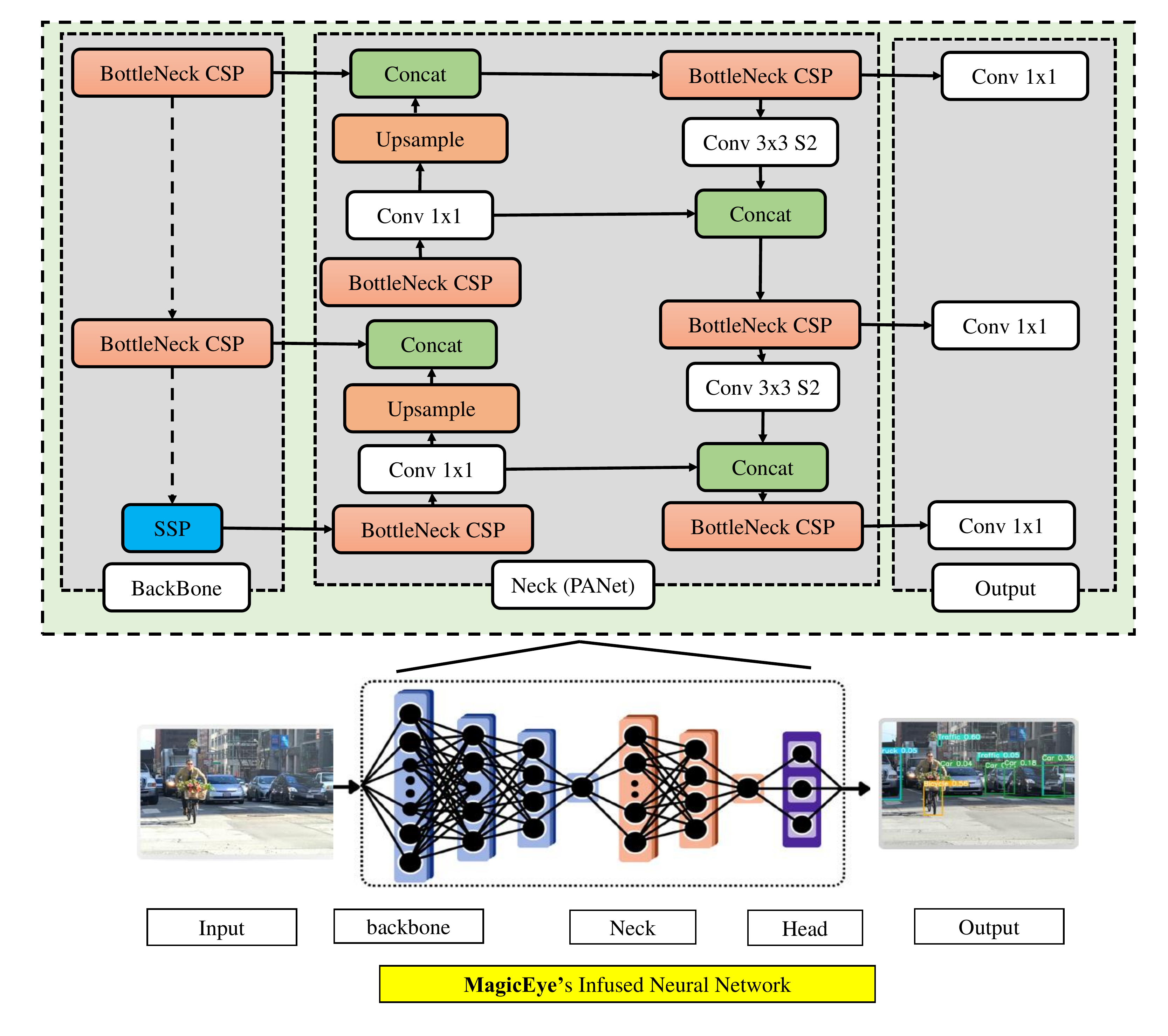}
\caption{An Architectural overview of the infused neural engine}
\label{FIG:model-arch}
\end{figure*}

The proposed MagicEye leverages the architecture of YoloV5 as its base model, thus adapting its performance and compactness \cite{redmon_you_2016}. In our task to develop a robust assistive system for the unsighted, speed and accuracy are highly critical, and the size of the back-end neural network determines the inference efficiency on resource-limited edge devices. To begin, our network incorporates the Cross Stage Partial Network (CSPNet), thus resulting in a reduced model \cite{wang_cspnet_2019}. Additionally, the inference speed and accuracy are significantly better due to incorporating gradient changes into the feature map. Furthermore, the base architecture operated Path Aggregation Network (PANet) \cite{liu_path_2018}, improving the information flow. Subsequently, the PANet improved the accuracy of object localization significantly. Finally, the Yolo layer, which acts as the head of our model, generates three different sizes of feature maps (20x20, 40x40, and 80x80) to achieve multi-scale prediction, allowing the model to handle small, medium, and more significant objects. Fig. \ref{FIG:model-arch} depicts the architectural overview of MagicEye's neural system.

YoloV5 has proven to be the state-of-the-art object detection algorithm and a superior version of its predecessors (YoloV1-V4). Numerous object detection algorithms have been proposed over the years, ranging from Sliding Window detectors to Region-based Convolutional Neural Networks. However, Yolo’s architecture has outperformed all previously proposed algorithms in terms of accuracy and speed while limiting the model's size \cite{redmon_yolov3_2018}. Yolo's architecture and unique working method enabled us to obtain significantly better results \cite{redmon_you_2016}. While previously developed systems could only detect and recognize a limited number of objects, MagicEye can detect and distinguish over 35 objects. As a result, MagicEye is not restricted to any particular environment and may thus be employed in any typical real-time scenarios.

\subsection{Facial Identification Network}

The most frequent object-identification activity carried out by sighted people is the identification of faces, which constitute a common group of visual objects. Faces stand out over other optical categories of objects, such as man-made objects, symbols, and numerals, since they are recognised nearly entirely by the sense of sight. However, it can be exceedingly difficult for visually impaired people to recognise faces. Therefore, the MagicEye is outfitted with cutting-edge facial identification models, allowing the device to recognise people immediately while ensuring high precision. The MagicEye, in particular, amalgamates two prominent face recognition architectures, namely, FaceNet \cite{schroff_facenet_2015} and VGG-Face \cite{qawaqneh_deep_2017}. Leveraging two disparate architectures allows the system to generate precise identification results while operating instantaneously.

\begin{figure}[htbp]
\centering
\includegraphics[width=0.9\textwidth]{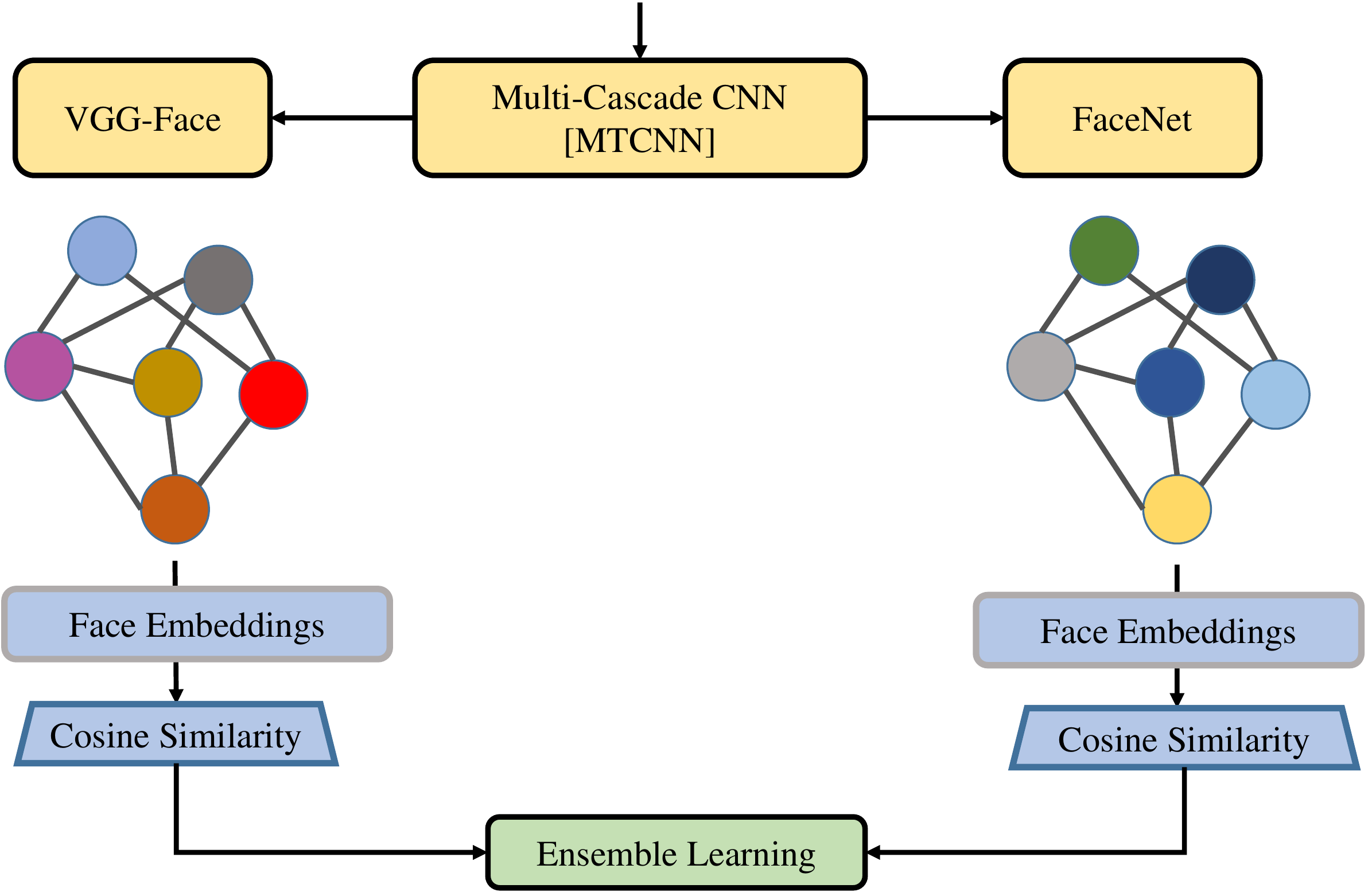}
\caption{Instancing of our ensemble facial recognition model.}
\label{FIG:Facial_recognition_model_architecture}
\end{figure}

The visually impaired individual would first register the faces of their preferred people. The model begins processing the identification automatically when the injected neural system recognises any faces in the environment, utilising the backend face detector, Multi-task Cascaded Convolutional Network (MTCNN) \cite{zhang_joint_2016}.  For every face, the neural network would eventually generate a set of embeddings and compare them to the user's database of registered faces. Utilizing the Cosine similarity function, as detailed in Equation \ref{EQN:cosine}, the similarity of faces would be precisely evaluated and verified, where $A$ and $B$ denote the face embeddings obtained from the feature extraction layer. An inclusive working of this module is elaborated in Fig. \ref{FIG:Facial_recognition_model_architecture}. Moreover, the integrated facial recognition model can further elaborate on other attributes such as gender, race and facial expression:
\begin{equation}
	Cosine Similarity = \dfrac {A \cdot B} {\left\| A\right\| \left\| B\right\| } 
	\label{EQN:cosine}
\end{equation}

\subsection{Currency Recognition System}

The majority of daily transactions that people conduct involve bank notes. Every person has to deal with cash daily for various reasons, including paying bills, renting, and other expenses. Unfortunately, performing hand-to-hand transactions proves challenging and arduous for an individual with a vision impairment. Due to repeated use, tactile markings on the surface of banknotes disappear or fade, making it difficult for those who are suffering with vision loss to detect and correctly identify banknotes by touch. The resolution to the underlying problem is found in the broad field of digital image processing, which involves extracting patterns and identification markers, and then comparing those to representations of actual banknotes. MagicEye adopts a CNN-based image classification model to identify between different denominations actively. Our suggested currency recognition model produced results with 94.5\% while using only 12810, culminating in a significantly efficient model. 

\subsection{Real-time working of MagicEye}

Utilizing MagicEye, the visually challenged can navigate unfamiliar environments without a problem. Our system allows the visually challenged to perform their daily chores effortlessly and prevents them from participating in any kind of accident. As stated earlier, our system is more complex than a basic object detection algorithm. 

To begin with the pipeline, the visually impaired individual would take an image of their surroundings using our proposed intelligent eyeglasses. The image will be captured once the user initiates the device by pushing a button on MagicEye's intelligent eyeglasses. Initially, the image will be processed by our integrated neural network, thus detecting and predicting all conceivable objects with their localization coordinates. The user can retrieve the generated results via auditory components such as Automatic Speech Recognition (ASR), Machine Translation (MT) and Text-to-Speech synthesis (TTS), allowing the individual to perceive the results through audio feedback and communicate with the device. Besides the user's manual trigger, MagicEye will be equipped with a proximity sensor that will alert the user and trigger the system when it encounters an obstacle nearby. 

As mentioned earlier, MagicEye additionally employs face and currency identification. The model automatically isolates the portions of the image and continues processing it for identification when the system detects any faces or banknotes. Specifically, the model generates face embedding for facial recognition by incorporating two disparate architectures. As a result of leveraging the results of two distinct models, our algorithm considerably reduced false positives. Aside from these functionalities, MagicEye includes a GPS to assist individuals in navigating streets and locations. By pairing the GPS and the audio device, the user will receive turn-by-turn directions to destinations.

\section{Experimental Results}
\label{Sec:Experimental_Results}

\begin{figure}[htbp]
\centering
\includegraphics[width=0.90\textwidth]{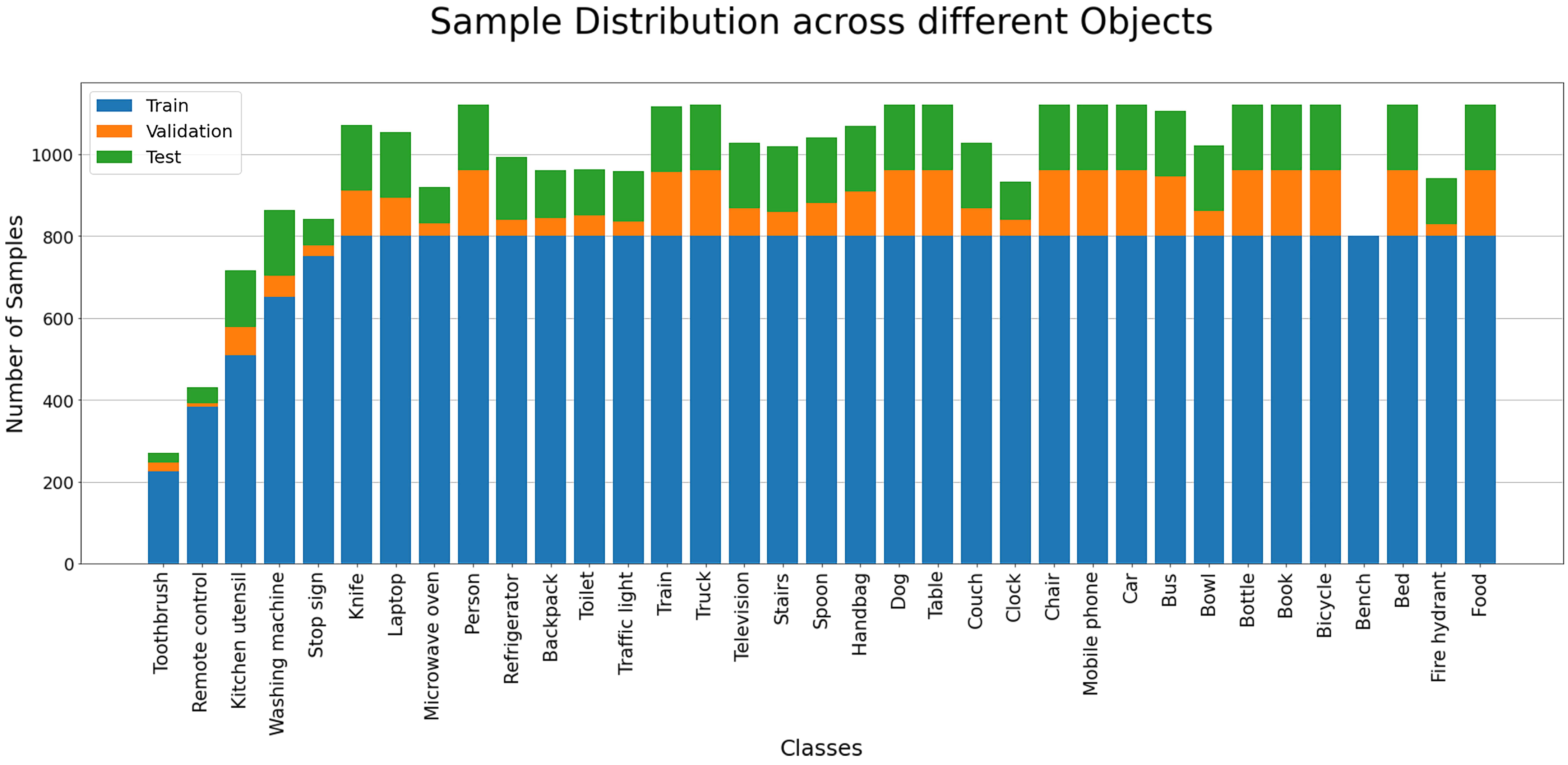}
\caption{Data split used to train, test, and validate MagicEye’s neural architecture.}
\label{FIG:Results_Data_split_Test-Train-Validation}
\end{figure}

One of the limitations of existing systems has been their ability to recognize a limited number of objects. Nevertheless, in our case, MagicEye is developed to detect and recognize 35 everyday objects encountered by a man in their day-to-day life. Notably, our inbuilt neural network is trained on a subset of the Open Images Dataset developed by Google \cite{kuznetsova_open_2020}. The Open Images repository contains over 600 boxable categories, from which we selected 35 relevant classes for our assistive system. The final training data set over 13,200 samples, each accompanied by a list of objects present in that image and their coordinates for localization. This enables the trained neural network to locate the objects and recognise them. Fig. \ref{FIG:Results_Data_split_Test-Train-Validation} illustrates the sample distribution among different classes.

\begin{figure*}[t]
\centering
	\includegraphics[width=0.98\textwidth]{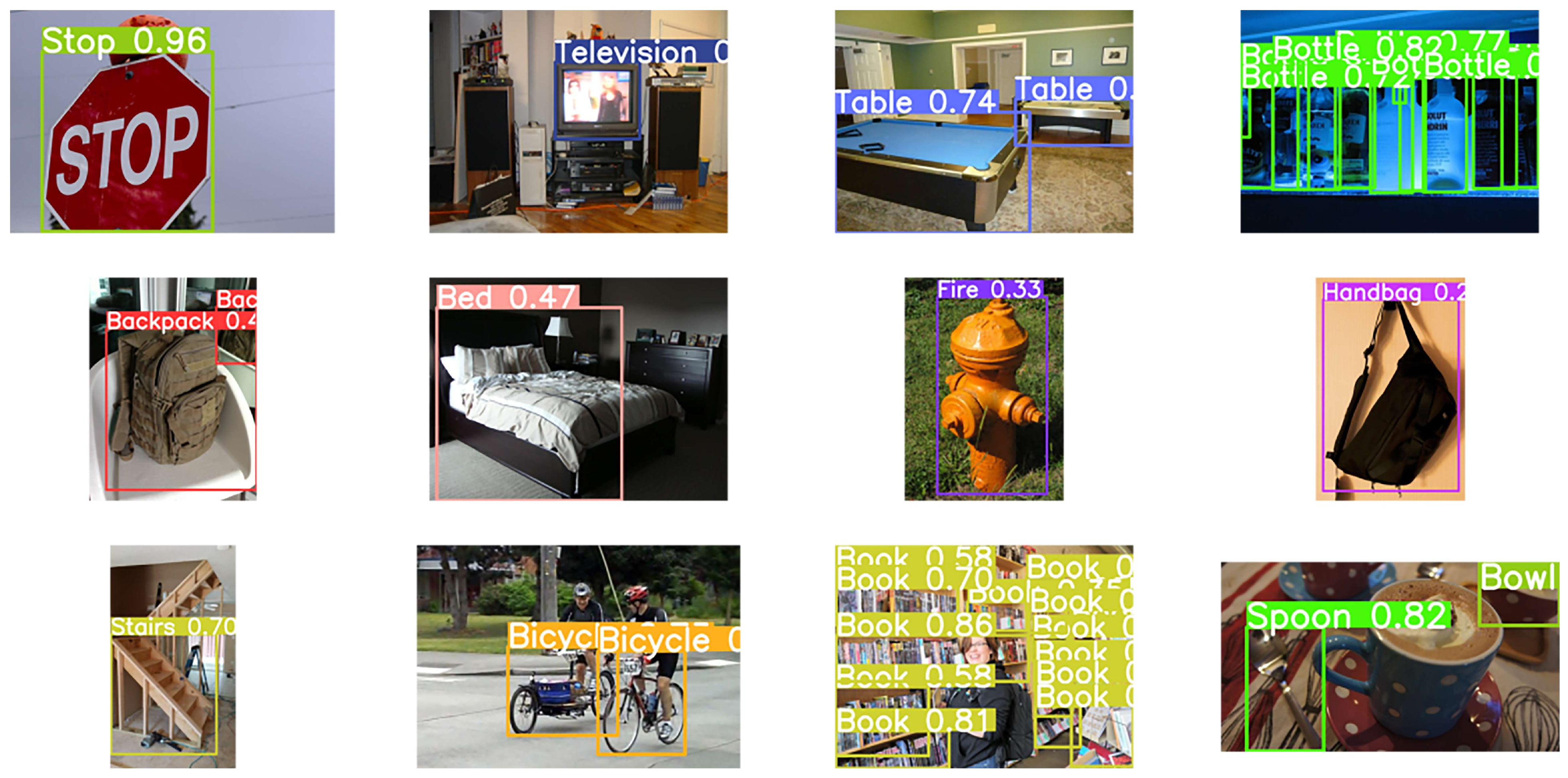}
\caption{Predictions generated by MagicEye on a few images from validation data split.}
\label{FIG:Results_MagicEye_Prediction}
\end{figure*}

\begin{figure*} [htbp]
\centering
	\includegraphics[width=0.95\textwidth]{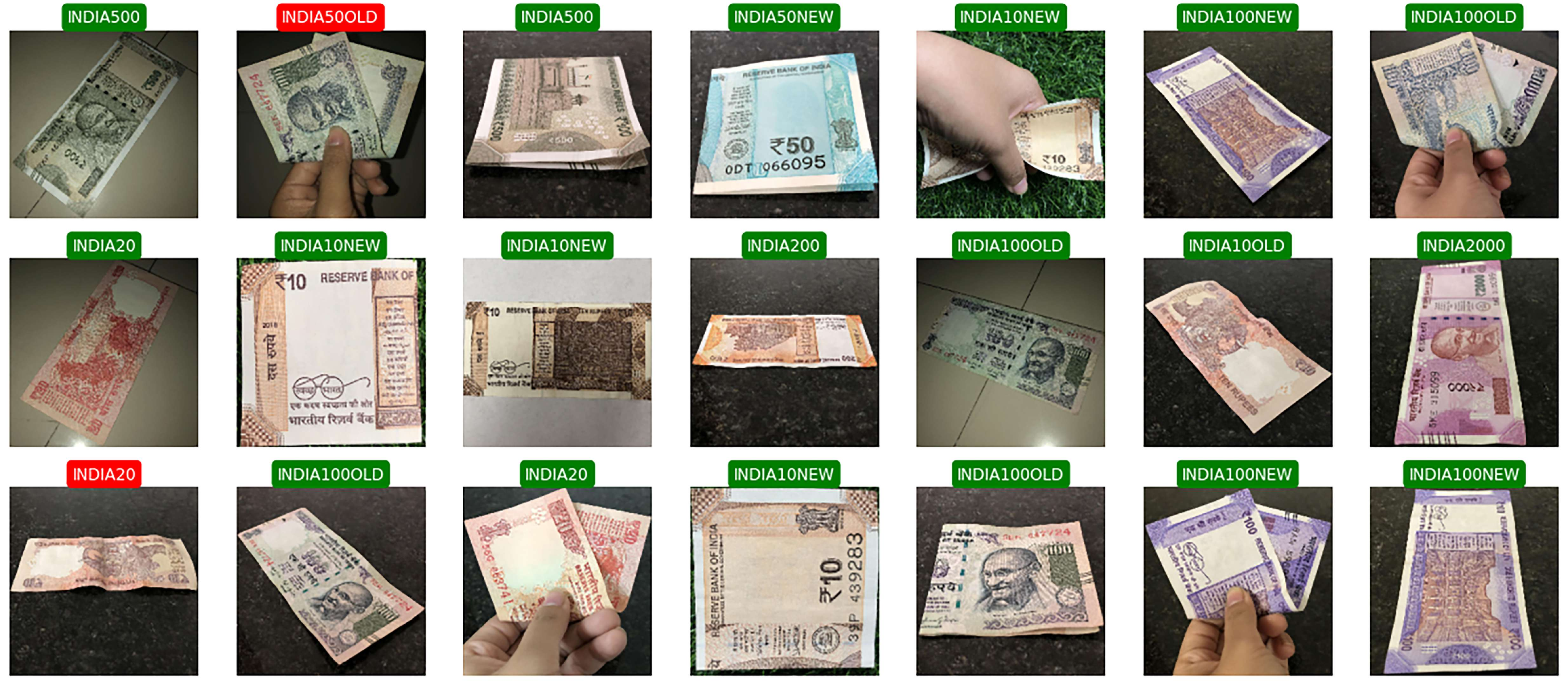}
\caption{Sample Predictions Generated by the Currency Identification Model.}
\label{FIG:Results_Currency_Prediction}
\end{figure*}

As previously stated, despite our model's capacity to recognize a wide range of objects and obstacles, our proposed methodology surpasses previously introduced systems in terms of accuracy, performance, and robustness. We evaluated our model on 1661 samples (roughly 10\% of the training data size) as part of the validation and testing procedure and obtained a mean Average Precision (mAP) score of 68.2\%. Compared to existing systems, our proposed methodology achieved a high mAP score and is capable of detecting more than 35 different classes. Fig. \ref{FIG:Results_MagicEye_Prediction} depicts a few predictions and outputs from our model. Furthermore, Fig. \ref{FIG:Results_Currency_Prediction} provides a visual representation of the sample predictions generated by the currency identification model, with green indicating a valid prediction and red representing an invalid prediction.

\subsection{Evaluation Criteria}

Several combinations of evaluation metrics are utilized to demonstrate the robustness of the proposed system which are typically used in the existing literature \cite{Joshi_TCE_2020, Sayeed_TCE_2019}.

\subsubsection{Mean Average Precision (mAP)}
The computer vision research community typically uses Mean Average Precision (mAP) as a benchmark metric to assess the stability of object identification models \cite{henderson_end--end_2017}. It compares the ground-truth bounding box to the detected box and outputs a score. Other assessment measures, including Confusion Matrix, Intersection over Union (IoU), Recall, and Precision, are often considered by the mAP score. Equation \ref{EQN:mAP} is used to compute the mAP score, where TP, TN, FP, and FN signify True Positive, True Negative, False Positive, and False Negative, respectively:
\begin{equation}
mAP\ =\ \frac{1}{n} \ \sum _{i=0}^{n} AP_{i}\\
\label{EQN:mAP}
\end{equation}

The mAP score is calculated by finding the Average Precision (AP) for each class and then calculating the mean of all classes. The mAP takes into account both false positives (FP) and false negatives (FN) and accounts for the trade-off between accuracy and recall (FN). Due to this characteristic, most detecting applications use mAP as a fundamental metric.

\subsubsection{Precision}
This metric assesses a model's ability to detect an object precisely. There are occasions when the model is triggered incorrectly and detects an object despite its absence in the image. For example, if the model detects 100 cars but only 90 are correct, the resulting precision would be 90\%. The precision is derived using Equation \ref{EQN:P}:
\begin{equation}
P\ =\left[ \ \frac{TP}{TP+FP} \ *\ 100\%\ \right]\\
\label{EQN:P}
\end{equation}

\subsubsection{Recall}
It is used to evaluate the model's performance in detecting objects based on the total number of objects in an image. For instance, if there are 100 cars in an image and the model can only recognize 75 of them, the recall is 75\%. The recall measures a model's ability to detect True Positives. Equation \ref{EQN:R} shows how the recall is calculated:
\begin{equation}
R\ =\left[ \ \frac{TP}{TP+FN} \ *\ 100\%\ \right]\\
\label{EQN:R}
\end{equation}

\subsubsection{Loss}
The Loss reflects the inaccuracies and flaws in the model. The more significant the loss, the more inaccurate the model. A neural network's training procedure iteratively evaluates the loss at each epoch, finding an optimal pair of weights and biases that will provide low loss. In particular, when training MagicEye's neural network, three losses are calculated for each epoch. Table. \ref{TBL:Loss-Data-Splits} depicts the obtained losses by our trained model.

\begin{table}[htbp]
\centering
\caption{Obtained losses by MagicEye on various data splits}
\label{TBL:Loss-Data-Splits}
\begin{tabular}{ p{2cm} p{2cm} p{2cm}}
 \hline \hline
  &Training &Testing\\
  \hline
Object Loss &0.0185	&0.0141 \\
Box Loss    &0.0286	&0.0327\\
Class Loss  &0.0063	&0.0155\\
\hline
\end{tabular}
\end{table}

Object loss represents the model's inability to detect an object, class loss quantifies the model's inaccuracy in predicting an object correctly, and box loss represents the model's inability to precisely identify the object centre while generating a bounding box that entirely covers the object.

\begin{figure*}[htbp]
\centering
	\includegraphics[width=0.95\textwidth]{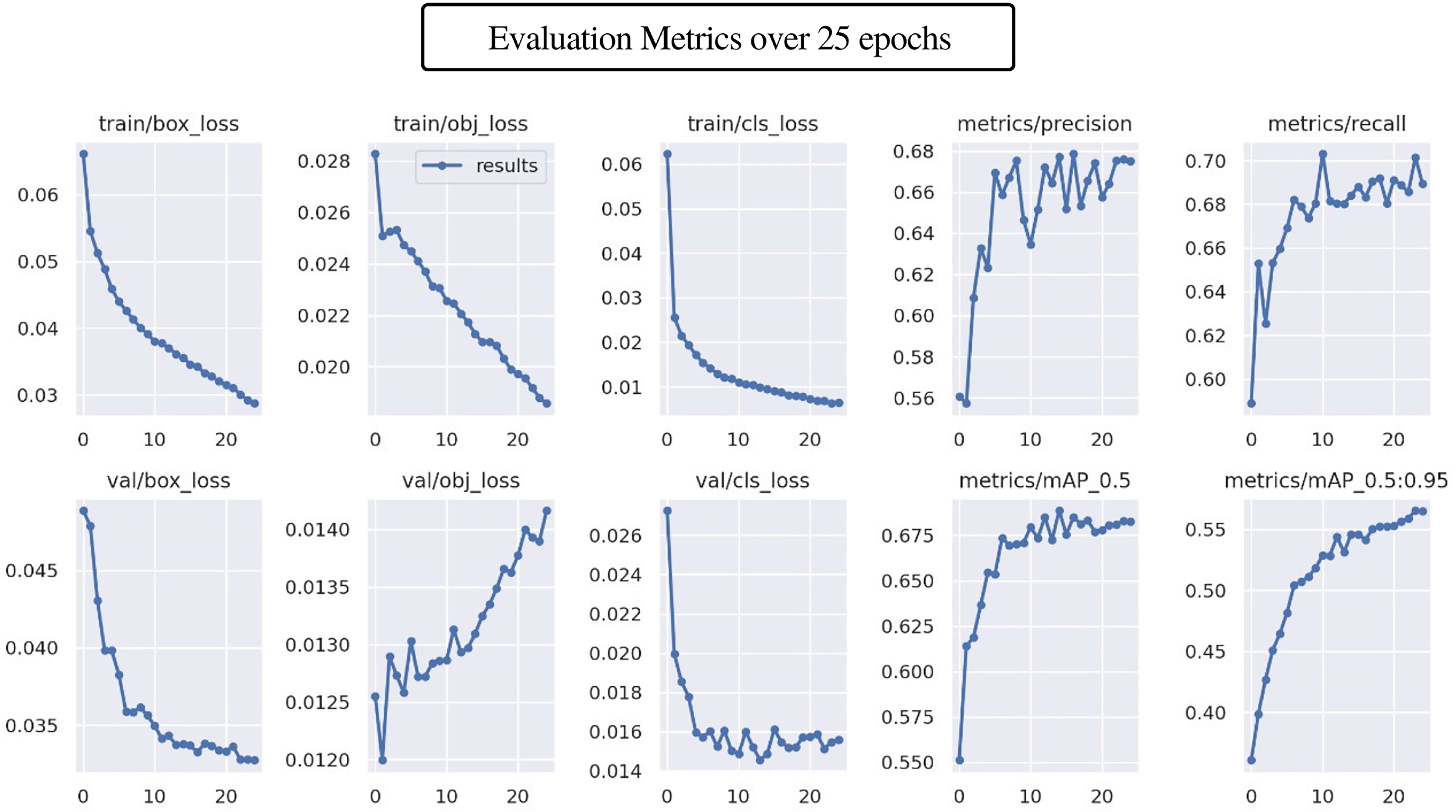}
\caption{Progression of evaluation metrics over 25 epochs.}
\label{FIG:Results_Evaluation_metrics}
\end{figure*}

\begin{figure}[t]
\centering
	\includegraphics[width=0.70\textwidth]{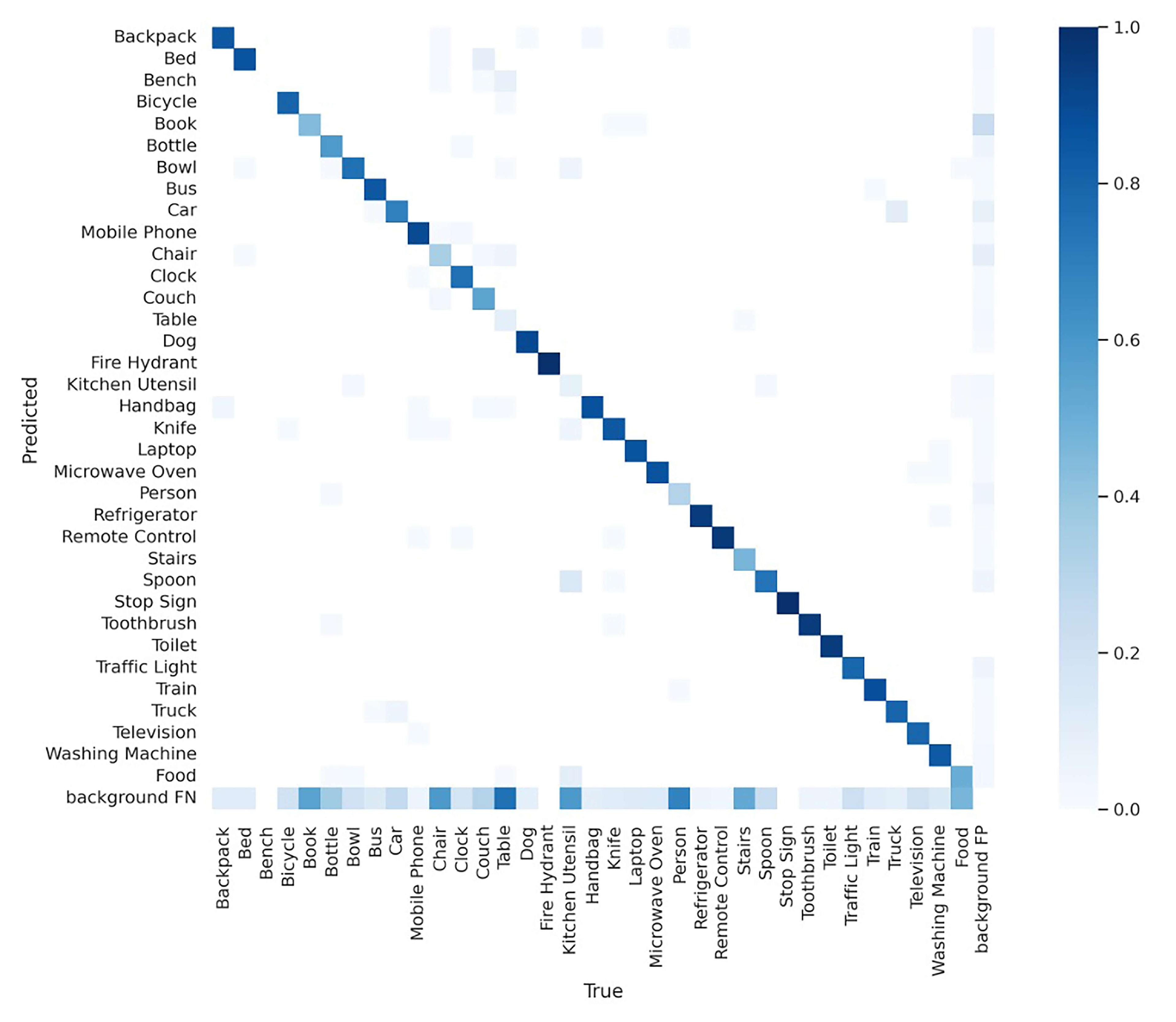}
\caption{Confusion Matrix generated by the neural model.}
\label{FIG:Results_Confusion_Matrix_MagicEye}
\end{figure}

\begin{figure}[htbp]
\centering
	\includegraphics[width=0.75\textwidth]{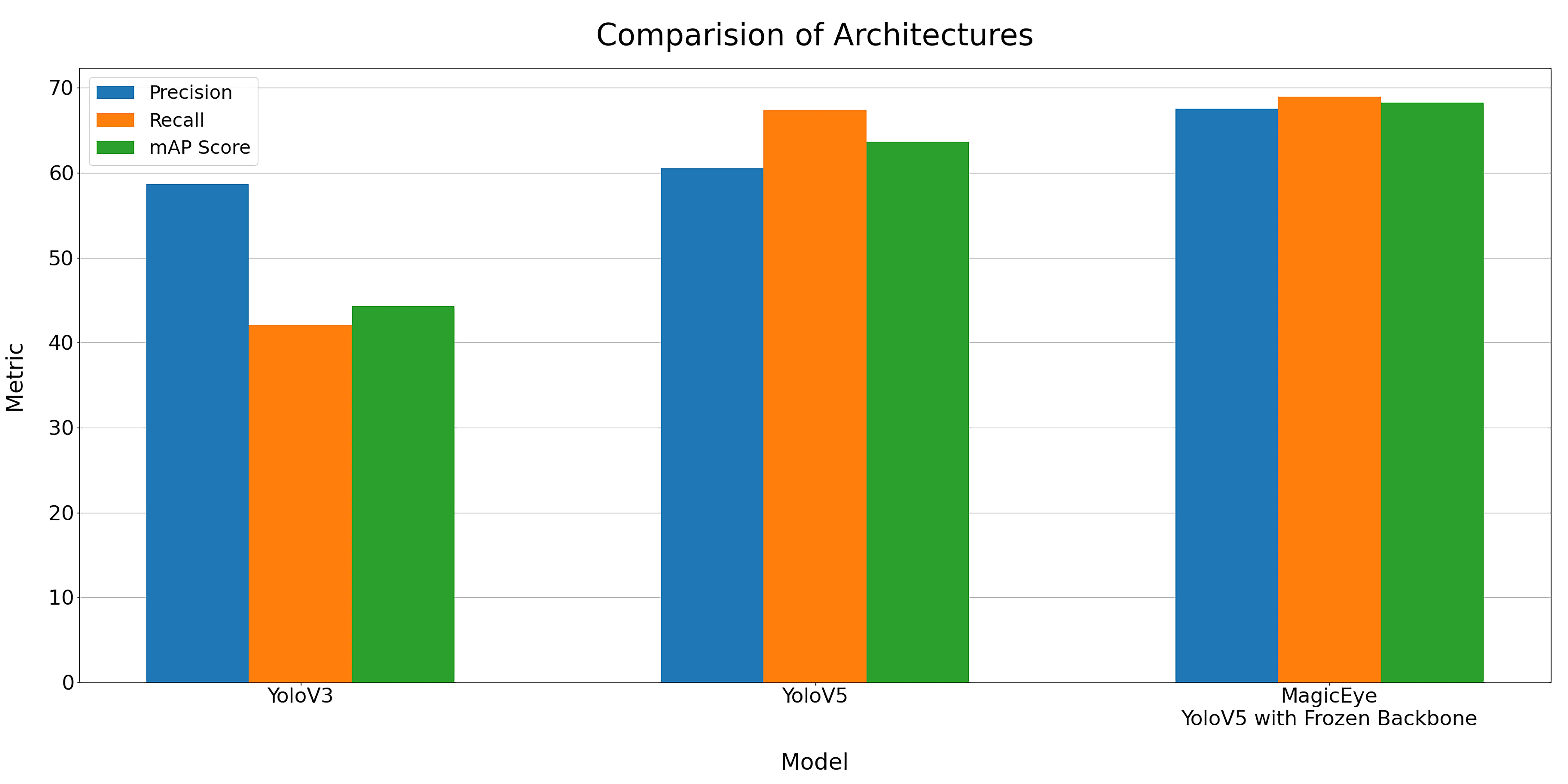}
\caption{Comparison of MagicEye with other state-of-the-art Yolo architectures.}
\label{FIG:Results_MagicEye_Comparative}
\end{figure}

Fig. \ref{FIG:Results_Evaluation_metrics} depicts the evolution of several metrics and loss over 25 epochs. It is evident that as epochs progressed, evaluation metrics, like recall, and precision, have progressively increased while the loss diminished. In addition,  a confusion matrix is employed to describe the model's performance on the test data. Each class categorizes the number of correct and wrong predictions produced by the model. The model's confusion matrix is depicted in Fig. \ref{FIG:Results_Confusion_Matrix_MagicEye}.

As previously indicated, we leveraged YoloV5's architecture to adapt its robust performance and precision. However, we compared our framework to other cutting-edge architecture, and MagicEye outperforms every comparison. MagicEye, in particular, employs YoloV5 architecture with a frozen backbone. By freezing weights, the gradient computations of the network are minimized, thus lowering the model's dimensionality. This increases efficiency by significantly reducing storage, transfer costs, and computation required for both the forward and backward passes. We compared our model to YoloV3 \cite{redmon_yolov3_2018} and a fully unfrozen YoloV5. The results are illustrated in Fig. \ref{FIG:Results_MagicEye_Comparative}.

In addition, a thorough evaluation was performed on the currency detection and facial recognition capabilities of MagicEye. As part of the evaluation process, we utilized various metrics such as precision, recall, and f1 score, as demonstrated in Table \ref{TBL:Model_results}. These metrics serve to assess the robustness of the model and provide a clear understanding of its performance. 


In conclusion, the MagicEye system represents a significant advancement in the field of visual aids for the visually impaired. Its real-time obstacle detection, turn-by-turn navigation, and advanced facial recognition and currency detection capabilities make it a comprehensive solution for the challenges faced by the visually impaired community. This research represents a crucial step towards making technology more accessible and inclusive for all individuals.

\begin{table}[htbp]
\centering
\caption{Obtained losses by MagicEye on various data splits}
\label{TBL:Model_results}
\begin{tabular}{ p{5cm} p{5cm} p{5cm}}
 \hline \hline \\
  Model &Currency Identification &Facial Recognition\\ \\
    \hline \\
  Dataset &Meshram et al. \cite{meshram_dataset_2021} &LFW \cite{LFWTech}\\
  Year &2021 &2007\\
  Accuracy &99.75\% &94.5\%\\
  F1-Score &99.86\% &94.21\%\\
  Recall &100\% &89.60\%\\
  Precision &99.72\% &99.34\%\\
  \hline
\hline
\end{tabular}
\end{table}



\section{Conclusion}
\label{Sec:Conclusion}

In this paper, we introduced MagicEye, an intelligent wearable device designed to assist the visually impaired with everyday tasks and navigation in both streets and indoors. Our proposed system is infused with a deep neural network that recognizes the surrounding items and obstacles and then communicates the results to the unsighted individual via auditory feedback. In addition, to assist individuals further, our proposed methodology incorporates facial and currency recognition modules. As a result, the user can detect and identify myriad faces and currency denominations in real time with high accuracy and precision. Moreover, our designed device features a GPS, which enables the user to receive turn-by-turn navigation instructions via aural feedback. Furthermore, the device comprises a proximity sensor that alerts the user when an obstruction is identified without any physical contact. When tested and evaluated against existing systems, our proposed framework yielded superior and remarkable results. In contrast to other proposed devices, MagicEye can identify various objects while maintaining high accuracy.

\section{Future Research}
\label{Sec:Future_Research}

By incorporating MagicEye into a real-world wearable device, we aim to provide users with an intuitive and seamless experience. Our next step is to adopt Active Learning and continually improve the device's recognition capabilities. This will enable the device to identify a wider range of objects and provide users with increasingly accurate and comprehensive feedback. In addition to enhancing the device's recognition capabilities, we also plan to integrate bone conduction technology to provide users with auditory feedback without blocking their ear canal. This will enable users to navigate their environment with greater ease and confidence. Moreover, this integration will enable users to receive feedback and navigate their environment simultaneously, which is crucial in complex and dynamic environments. We envisioned that unified solutions which combine mechanisms for automatic stress detection, glucose level detection, and fall detection in the healthcare cyber-physical system (H-CPS) can improve quality of live significantly \cite{Rachakonda_TCE_2021, Rachakonda_IFIP-IoT_2021, Joshi_TCE_2020, Sayeed_TCE_2019}.

\bibliographystyle{IEEEtran}
\bibliography{Bibliography_MagicEye}


\begin{IEEEbiography}
[{\includegraphics[height=1.2in,keepaspectratio]{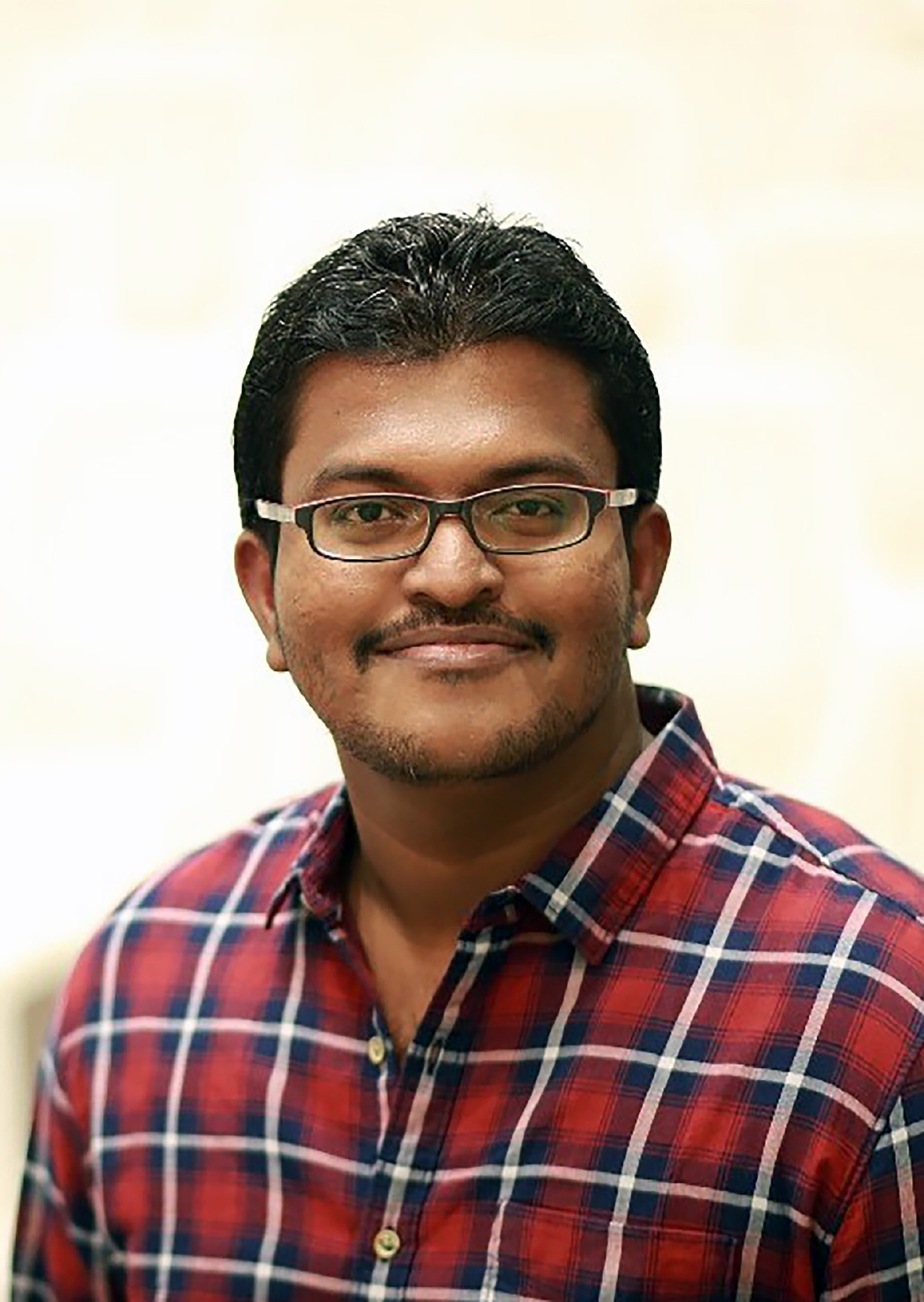}}]
{Sibi C. Sethuraman} (M'18) received his Ph.D from Anna University in the year 2018. He is an Associate Professor in the School of Computer Science and Engineering at Vellore Institute of Technology – Andhra Pradesh (VIT-AP) University. 
He was an Assistant Professor in the Department of Computer Science and Engineering at VIT-AP from May 2018 to September 2018. 
He is a visiting professor and a member of Artificial Intelligence Lab at University Systems of New Hampshire (KSC campus) in the Department of Computer Science. 
Further, he is the coordinator for Artificial Intelligence and Robotics (AIR) Center at VIT-AP. 
He is the lead engineer for the project VISU, an advanced 3d printed humanoid robot developed by VIT-AP. 
He is an active contributor of open source community. 
He is an active reviewer in many reputed journals of IEEE, Springer, and Elsevier. He is a recipient of DST fellowship.
\end{IEEEbiography}

\begin{IEEEbiography}
[{\includegraphics[height=1.2in,clip,keepaspectratio]{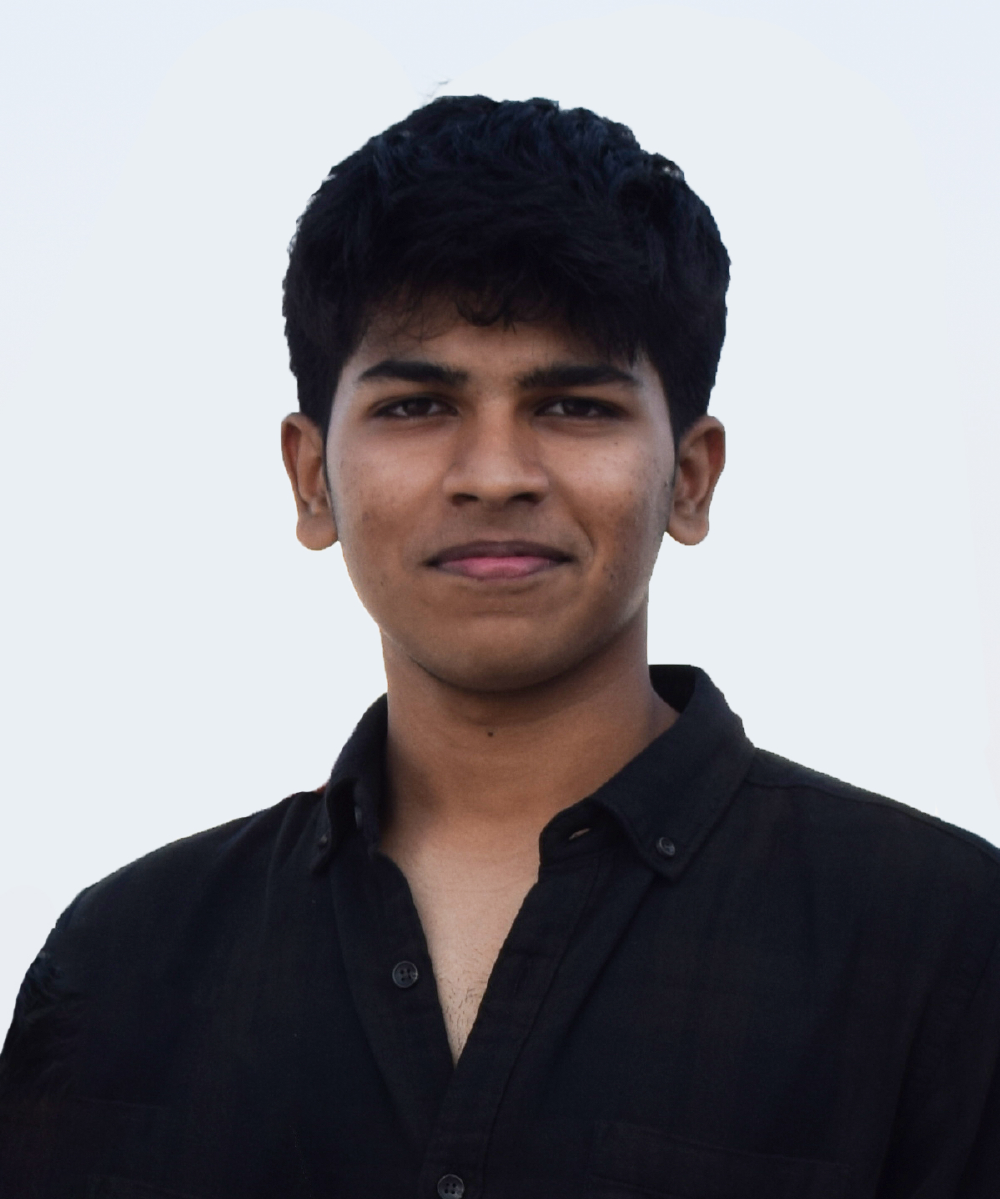}}]
{Gaurav Reddy Tadkapally} (S'19) is a Bachelor of Technology student at Vellore Institute of Technology, Amaravati (VIT-AP). He specializes in Deep Learning and the applications of Artificial Intelligence in Consumer Electronics and Embedded Hardware. As a member of the Artificial Intelligence and Robotics center at VIT-AP, Gaurav has developed several cutting-edge technologies, including iDrone, an IoT-powered drone for detecting wildfires, and SimplyMime, a hand gesture-based remote control alternative.
\end{IEEEbiography}

\begin{IEEEbiography}
[{\includegraphics[height=1.25in, keepaspectratio]{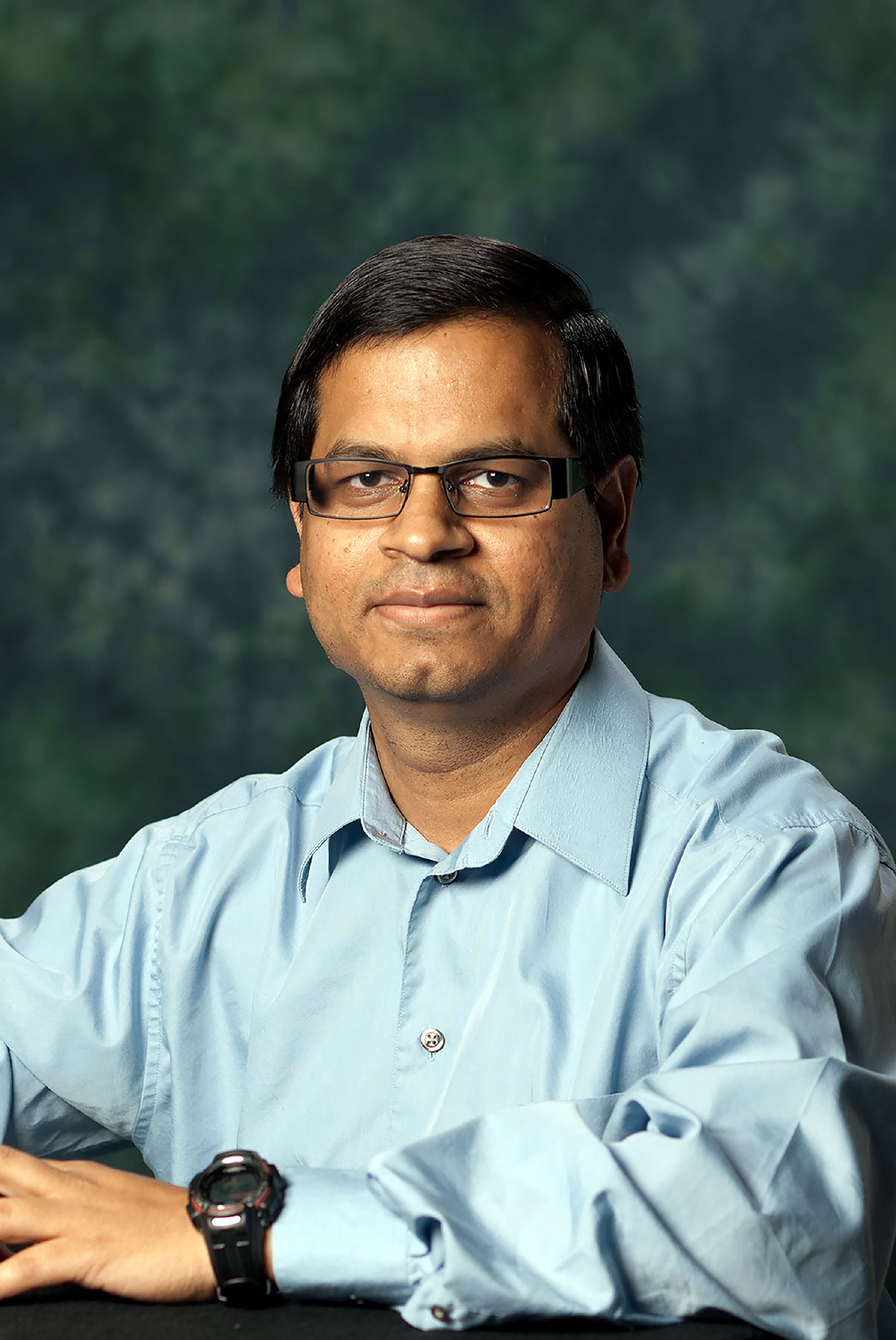}}]
{Saraju P. Mohanty} (Senior Member, IEEE) received the bachelor's degree (Honors) in electrical engineering from the Orissa University of Agriculture and Technology, Bhubaneswar, in 1995, the master's degree in Systems Science and Automation from the Indian Institute of Science, Bengaluru, in 1999, and the Ph.D. degree in Computer Science and Engineering from the University of South Florida, Tampa, in 2003. He is a Professor with the University of North Texas. His research is in ``Smart Electronic Systems'' which has been funded by National Science Foundations (NSF), Semiconductor Research Corporation (SRC), U.S. Air Force, IUSSTF, and Mission Innovation. He has authored 450 research articles, 5 books, and 9 granted and pending patents. His Google Scholar h-index is 49 and i10-index is 211 with 11,000 citations. He is regarded as a visionary researcher on Smart Cities technology in which his research deals with security and energy aware, and AI/ML-integrated smart components. He introduced the Secure Digital Camera (SDC) in 2004 with built-in security features designed using Hardware Assisted Security (HAS) or Security by Design (SbD) principle. He is widely credited as the designer for the first digital watermarking chip in 2004 and first the low-power digital watermarking chip in 2006. He is a recipient of 16 best paper awards, Fulbright Specialist Award in 2020, IEEE Consumer Electronics Society Outstanding Service Award in 2020, the IEEE-CS-TCVLSI Distinguished Leadership Award in 2018, and the PROSE Award for Best Textbook in Physical Sciences and Mathematics category in 2016. He has delivered 18 keynotes and served on 14 panels at various International Conferences. He has been serving on the editorial board of several peer-reviewed international transactions/journals, including IEEE Transactions on Big Data (TBD), IEEE Transactions on Computer-Aided Design of Integrated Circuits and Systems (TCAD), IEEE Transactions on Consumer Electronics (TCE), and ACM Journal on Emerging Technologies in Computing Systems (JETC). He has been the Editor-in-Chief (EiC) of the IEEE Consumer Electronics Magazine (MCE) during 2016-2021. He served as the Chair of Technical Committee on Very Large Scale Integration (TCVLSI), IEEE Computer Society (IEEE-CS) during 2014-2018 and on the Board of Governors of the IEEE Consumer Electronics Society during 2019-2021. He serves on the steering, organizing, and program committees of several international conferences. He is the steering committee chair/vice-chair for the IEEE International Symposium on Smart Electronic Systems (IEEE-iSES), the IEEE-CS Symposium on VLSI (ISVLSI), and the OITS International Conference on Information Technology (OCIT). He has mentored 2 post-doctoral researchers, and supervised 14 Ph.D. dissertations, 26 M.S. theses, and 18 undergraduate projects.
\end{IEEEbiography}

\begin{IEEEbiography}
[{\includegraphics[height=1.2in,clip,keepaspectratio]{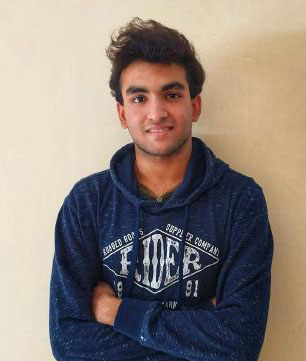}}]
{Gautam Galada} (S'22) is pursuing a Bachelor of Science in Computer Science with a specialization in Artificial Intelligence at Vellore Institute of Technology, Amaravati (VIT-AP). His areas of focus include Deep Learning, Internet of Things, and Machine Learning. He is the Co-Founder of Digital Fortress, established at VIT-AP, and currently serves as the Lead Researcher at the Center of Excellence for Artificial Intelligence and Robotics.
\end{IEEEbiography}

\begin{IEEEbiography}
[{\includegraphics[height=1.2in,clip,keepaspectratio]{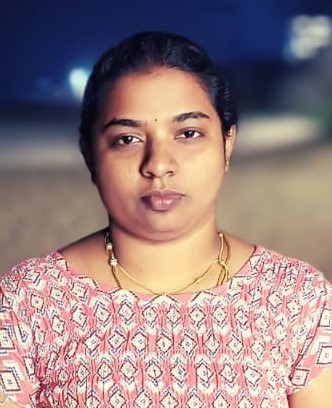}}]
{Anitha Subramanian} (M'22) received the bachelor's degree in Electronics \& Communication Engg. from the Anna University, India, in 2017, the master’s degree in Power Electronics from the Anna University University in 2019, and she is currently pursuing her Ph.D. degree in Electronics Engineering from VIT-AP University.
\end{IEEEbiography}

\end{document}